# Cultivating Software Performance in Cloud Computing

Li Chen, Colin Cunningham, Pooja Jain, Chenggang Qin and Kingsum Chow

{li.chen, cunningham.colin, pooja.jain}@intel.com, {chenggang.qcg, kingsum.kc@alibaba-inc.com

## Abstract

There exist multitudes of cloud performance metrics, including workload performance, application placement, software/hardware optimization, scalability, capacity, reliability, agility and so on. In this paper, we consider jointly optimizing the performance of the software applications in the cloud. The challenges lie in bringing a diversity of raw data into tidy data format, unifying performance data from multiple systems based on timestamps, and assessing the quality of the processed performance data.  Even after verifying the quality of cloud performance data, additional challenges block optimizing cloud computing. In this paper, we identify the challenges of cloud computing from the perspectives of computing environment, data collection, performance analytics and production environment.

## 1   Introduction

Cloud computing has become increasingly important. Deploying diverse applications upon cloud infrastructure introduces layers of complexity. To gain understanding and design a more efficient cloud, system and software performance data over compute, memory, storage and network are collected from virtual and physical machines. Examining these immense cloud performance datasets and uncovering patterns presents a daunting challenge. The model of computation is such that hundreds of compute nodes host thousands of virtual machines to deliver a diverse range of software over the internet. The cloud meets storage needs with a combination of block and object storage interacting with application and database virtual machines to provide data management functionality. Cloud management requires a set of administration nodes and hypervisors distributed on physical compute server resources to orchestrate the cloud.

To gain understanding and propose more efficient cloud management solutions, performance data is collected from virtual machines, administration nodes, computing nodes, and so on. Unlike enterprise applications, cloud performance optimization has many aspects under consideration. First, from a macro perspective, multiple applications in the cloud data center are running at the same time. Optimizing for a single software application is likely to prove suboptimal for the entire cloud computing environment. Second, applications running on virtual machines are tenants on the same computing nodes, and consequently share memory bandwidth, cache, and network. Mixing of workloads and using the same shared resources causes the problem of noisy neighbors and performance degradation. Third, from a micro-scope perspective, characterizing individual applications running on each virtual machine or individual applications running on multiple virtual machines requires processing and analyzing performance from multiple systems. Within an individual application, analyzing the correlation of events happening on one system versus another adds another layer of complexity. Last but not least, with the complicated relationship between all the components in the cloud, making a rapid data-driven cloud management decision is almost impossible without processing and deriving useful information from a large cloud performance dataset. To meet industry standards for providing a rapid and efficient management decision, a data-driven approach to deliver immediate analysis of the cloud and providing analytical intelligence of the cloud metrics is needed. Examining large cloud datasets and uncovering hidden patterns is not easy. Yet, we hope to influence the cloud computing industry should it be able to discover such patterns, identify solutions to satisfy user experience and translate these learnings into blueprints that enable more efficient cloud management solutions.

In [1], we describe a process, implemented in software, to assess the quality of cloud performance data. This process combines performance data from multiple machines, spanning across user experience data, workload performance metrics, and readily available system performance data. The data collection and processing procedure across the system is able to generate concrete data for posterior analytics. Principled statistical analysis on cloud performance data will serve as a valuable tool to assess cloud performance. One challenge of dealing with performance data sets is bringing them into a tidy data format [10] for analysis. In [1], the process we proposed addresses this challenge and prepares the collected cloud performance data readily for data-driven analytics.

In this paper, we discuss additional challenges for cloud computing. We describe these challenges from the aspects of computing environment, data collection, performance analytics and production environment. The value of understanding the challenges helps us seek better solutions to improve cloud performance analysis.

In Section 2, we review different types of cloud performance data and how our previous process [1] enables cloud performance data cleansing. In Section 3, we discuss the challenges for cloud computing from the analytics perspective and production perspective. Section 4 summaries our paper and discusses future directions for this area of investigation.

## 2  Cloud Data Diversity

In this section, we describe the diverse range of cloud performance data, acknowledge the challenges for processing the data, and review our proposed approach [1] on assessing the quality of cloud performance data.

User experience data, such as throughput and response time, can be obtained from load driver systems typically used in software testing. One such load driver, Faban [4], is a driver development framework used in SPECjEnterprise [5], SPECvirt [6] and SPECsip [7] benchmarks, and can control a number of load generation nodes. Such a framework defines operations, transactions and associated statistics collection and reporting. Faban is one of the tools that can be used for cloud computing workloads.

System Activity Report (SAR) [8] on Linux systems can save system performance data such as CPU activity, memory/paging, device load, network, etc. SAR logs the contents of selected cumulative activity counters in the operating system. The accounting system, based on the values in the count and interval parameters, writes information a specified number of times spaced at the specified intervals in seconds. Performance Counters for Linux (Perf) [9], a Linux profiling tool for performance counters, can add additional detailed performance counters for software processes and microarchitectures.

### 2.1  The challenges in preparing cloud performance data

One challenge of dealing with performance data sets is bringing them into a tidy data format [10] for analysis. Before discussing these challenges, we first offer motivation for why such an action is desired. Cloud workloads are much more complex than enterprise workloads. Manually examining the cloud performance data is not feasible. The workload data must be restructured to allow statistical modelling. Dynamic analysis on time series cloud performance data provides new insights for cloud performance optimization. Techniques such as model selection, non-stationary time series analysis, and stochastic processes can be adopted to analyze the cloud workload. Furthermore, one could also borrow the techniques such as spectral clustering [11], sparse representation classification [12] [13], vertex nomination [14] [15] and graph matching [16] [17] in the field of random graph inference to model the user behavior graph of the cloud workload. All these inference frameworks and methodologies would benefit from a tidy data format. Only then can data analysis yield insights on the dynamics of cloud performance. As different performance tools generate data in different formats, parsing the data into a simple and coherent format is just as essential as collecting and analyzing the data. Moreover, validating the quality of the data processing procedure is necessary; otherwise all subsequent work can be suspect. In this section, we first start with a primitive and idealized scenario for quality assessment the data processing. We then define the challenges in identifying performance bottlenecks in the cloud, and present our insights on dealing with and working around these challenges.

There exists no standard format in which the data are collected. Still, we posit that parsing the data into a simple and coherent format is equal in importance to the effort involved in collecting and analyzing the data. Data munging gets tedious when dealing with data sets collected from different sources. Merging these files across the date and time spectrum is a particular challenge. Elaborating on this, various data sets might have different formats of time in which the data are collected, namely the a) world clock time and b) epoch (UNIX time), which might further differ in the following criterion: i) time zones, ii) units of measurements (seconds or milliseconds), iii) the time interval at which the data is collected and iv) the frequency at which data samples are collected. Cloud performance data sets are simply too huge to manually check consistency of results produced by the software components.

## 2.2 Review of our previously proposed approach

Here we present a brief review of our previously proposed approach [1] for assessing the quality of cloud performance data. To overcome these challenges, our approach incorporates multiple layers of quality assessments. Our goal is to assess the quality of the processed data for further analysis. In a nutshell, we first add a set of basic queries and scripts to check the correctness. As data can vary greatly, we then add a set of performance models to check the quality of the performance assessment. By using ensembles of performance models, we minimize potential errors in the software performance assessment tool chains. Figure 1 depicts the flow of our proposed software. In the following subsections, we describe in detail how our software works.

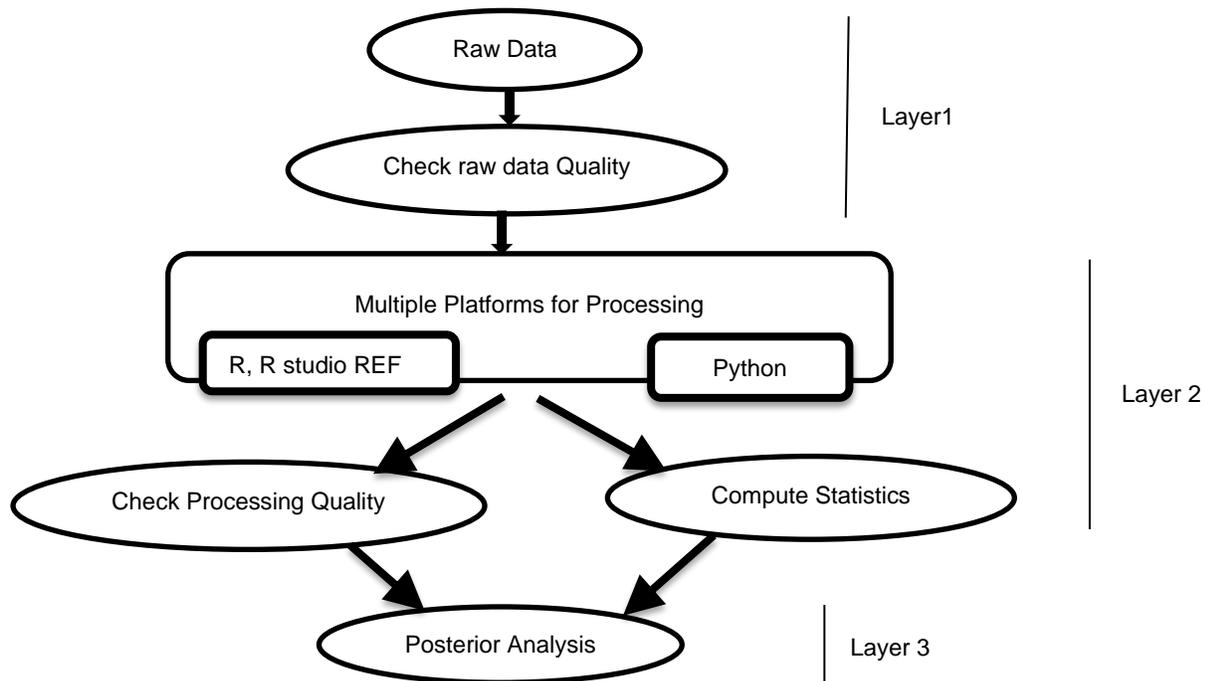

**Figure 1. A quality model for Cloud performance analytics**

# 3 Identifying the Challenges for Cloud Performance

## 3.1 Challenges of Cloud Computing Performance Guarantee

Cloud computing offers the promise to efficiently share resources among tenants and their workloads but verifying this aspiration and searching for improvement and optimization opportunities for the cloud administrator is not simple. The resources we first focus upon might be categorized into compute, memory, network and storage. In the case of Infrastructure-as-a-Service (IaaS), the cloud service provider may dedicate an amount of virtual CPUs (vCPUs) along with access to a fixed amount of physical memory and temporary working disk, based on the customer service request. Storage needs are often serviced a la carte.

One might think that the pure compute resources are immune to noisy neighbors but today's virtualized infrastructure employ multi-core processors potentially execute multiple threads per core and allow the hypervisor great flexibility in provisioning VCPUs. Many cloud service providers (CSPs) use Intel Xeon CPUs with Hyper-Threading Technology which allows two threads to process simultaneously. To the operating system each processor core has two virtual CPUs (vCPUs). The hypervisor can create a virtual machine (VM) with a configurable number of VCPUs. Because many VMs are often placed on a single server, the VCPUs for a single VM will often reside next to another instance or perhaps none at all. The cloud performance analyst must be cognizant of the VCPU placement and adjacent VM and VCPU activity as compute resources on any individual VCPU can flex higher if the potentially adjacent VCPU is absent or idle and available. This complexity will manifest in slightly variable compute resources to the VM one level up.

For each physical server, the memory subsystem is shared among administrative and customer instances with some complexity. By default, memory cache and memory bandwidth are shared among virtual machines within a socket. Workloads differ greatly in the cache capacity required and the frequency of physical memory access. As such, VM instances memory caching and bandwidth requirements dynamically modulate memory resources available to neighboring instances while themselves being impacted by neighbor demands. We believe that being able to monitor and perhaps control memory activity will yield understandings that can positively influence cloud orchestration and VM density by way of smarter VM placement and memory control. It should also yield insight into the system memory requirement most appropriate to workload types.

The computing instance also attaches generally to networking services that are shared by other instances at various levels as you move up in the network hierarchy from the NIC to the initial switch and onward. Networking is typically less described to the customer and is most often shared among other instances on the server, within the data center and beyond onto the Internet. As to the network, a primary concern for the cloud operator to ensure that the network does not impede the quality of service or degrade the customer experience. But, to know if and when the network is a bottleneck to workload performance, and to understand where in the network the constraint lies, a rich and thoughtful data collection across the data center is required. The cloud analyst must be aware of the network topology and its capabilities. We may then instrument the monitoring of the local virtual and physical NICs as well as individual and aggregate use of switches throughout the datacenter.

Because cloud storage is very often delivered over the network, storage is not immune to noisy neighbor effects. A VMs I/O performance is subject first to all the networking forces mentioned above. The I/O devices a VM sees are likely to be virtual devices that route to a physical device over the network. The cloud analyst must be able to map physical devices to the VMs in use. Physical devices space will also be shared among users so concurrent use of the physical device also injects noisy neighbor concerns.

We should not close this section before bringing attention to another shared infrastructure common to virtualized systems are those resources set aside or used by the virtual machine manager and the hypervisor. It is common for a CSP to reserve a fixed number of VCPUs per physical server for local virtualization management purposes. These VCPUs may well be pinned to a set of physical cores perhaps not available for customer subscription. Clearly a CSP would want to minimize the non-revenue compute available yet provide enough resources for local orchestration and execution. In the case of Xen, the initial domain, known as Dom0 is a privileged domain that starts and manages those hosted on the server. Cloud operations will need to monitor the compute, memory, networking and I/O use on this VM instance along with the customer VMs. This is especially important when the burden of hosted VM networking or I/O puts demand on compute processing or if networking is transparent to customer VM as might be the case when for I/O (e.g., for paravirtualized guests in Xen 4.3). Monitoring here can help cloud operations as they strive to expose the indirect resource demands the customers place in order to improve both billing and architecture practices.

### 3.2 Challenges of Collecting Performance Data in a Cloud environment

Collecting performance data in a distributed environment is challenging in many ways, a few of which we will highlight in this section. Firstly, in a cloud environment, we have number of virtual machines residing on the same or different hosts talking to each other over a common network.

Synchronous data collection in such environments is very difficult. It would require configuring and deploying a complex software application which would enable synchronized automation of performance monitoring tools on all the machines. Enabling of such software systems requires automation of secure communication in the cluster. Setting up secure connections over networks involves much tedious effort like creating secure keys, users, and links, unique subnets and IP address, bridges on VMs, etc. Even one error in the setup can break the connection and will result in incomplete and inaccurate data.

Secondly, the version of the tools we use have a lot of dependencies on the operating systems, kernel, hardware specifications etc. So to call out in simple words, selecting a single performance monitoring tool that can collect all the data we need from all machines in our cluster could be challenging. For example, the open source performance monitoring tools, like sar and perf are exclusive to UNIX/LINUX-based operating systems, and not for WINDOWS. This compels us to use multiple performance monitoring tools, adding to the complexity of the VM environment before starting performance monitoring tools.

The real challenge comes in the phase of data interpretation. Every tool, has its own way of computing results—keeping aside the data formats (time notation, units, frequencies, etc.), the way in which each value/data point is calculated could vary vastly from tool to tool. By example, suppose tools A, B and C report data every 10 seconds on three different machines performing inter-dependent functions, Tool A takes an average of the data observed in the last 10 seconds and gives a value, while tool B, reports whatever it measures at 10$^{th}$ second and Tool C reports data that it measured for the 5th second of the 10s interval. Now when we put the data together after an experiment, in a time-series sheet to analyze the performance and resource utilization of our machines during our experiment, the data could not support consistent analysis or comparison.

The right approach would be to take in account the internal computation of every counter of every tool we use to monitor data and adjust our data accordingly before doing further analysis. This in itself is a huge challenge and lot of hard work and planning.

### 3.3 Challenges of Analytics on Cloud Performance Data

The cloud performance data include both event and sample-based data logs. Putting the time series of performance data back to the same scale becomes challenging. Even though the sampling intervals for different performance monitoring tools may be the same, the actual timestamp for each type of performance data might not lie on the exact same timestamp as another. Therefore, by merging the time series data as described in [2], a lot of missing values introduced for unmatched points in time. The presence of missing data can cause challenges for correlation analysis, because the mathematical formulation requires each variable hold a value for each observation to be included in the correlation. We have found that appropriate missing value imputation can address this challenge and comes to rival in importance any subsequent statistical analysis methodology. There is a vast of literature describing different techniques of missing value imputation. For instance, one can choose to impute the missing values by the mean, median, maximum or minimum. Or one can also impute the missing values by interpolation over the previous and subsequent observed values. In cloud performance analysis, the choice of missing value imputation techniques depends on the performance metrics. For example, garbage collection (GC) are event-based and occurs when resources need to be reclaimed, while system activity report (SAR) data are sample-based and occurs at fixed length of time intervals. When merging garbage collection logs and system activity report data, blocks of missing values are introduced. However imputing the missing values for different performance metrics requires contextual knowledge. When imputing the heap size of GC logs, imputing the missing values by zero makes sense, because at the missing timestamps, no GC activity occurred. On the other hand, one could use linear interpolation for imputing many SAR performance metrics. If taken during the steady state of a workload, one could use the mean or median to impute the missing values for SAR performance metrics. It is indeed a challenge to identify the optimal missing value imputation techniques for performance analysis.

Another challenge for drawing inference on merged performance data lies in data transformation. The need for data transformation becomes essential when we want to spot patterns and detect changes across performance metric types. CPU utilization lies between 0-100%, network activity is measured by kilobytes per second often the order of $10^6$, while cache misses per thousand instructions could be in the order of $10^{-3}$. Plotting such metrics on a common scale cannot expose the behavior changes, because the very low valued performance metrics will display as almost a straight line. There are many ways of data transformation such as centering, data normalization, log transformation, inverse transformation, etc. Different types of data transformation will illuminate (or potentially mask) different patterns of behavior changes. Furthermore, it will result in a different result of correlation analysis. Therefore, providing a principled methodology of transformation is a necessary if difficult task.

The complexity of cloud computing environment is beyond manual inspection. When merging cloud performance data, the number of performance metrics, or statistically speaking, the number of variables, is usually in the thousands. Manually inspecting which performance metrics are more relevant to each other is virtually impossible; to infer causal relationship among the performance metrics is doubly so. Therefore, we find examination of correlation analysis results is an efficient way in the early journey to identify relevant metrics. Directly calculating the correlation between two performance metrics is not sufficient, because the data has inherent temporal effects. We propose using cross-correlation analysis, which considers the time lagging effect to assess the correlation between performance metrics. Although this approach is more suitable to identify correlated time series, it is challenging to determine how wide or narrow the time windows should be for identify the temporal correlation.

# 4  Related Work

Bianque is a software quality performance monitoring system deployed in Alibaba. It enables performance data collection from thousands of servers. As such, it can be viewed as a software quality performance analysis platform. To aid software quality analysis, it can extract performance profiles from the servers. From those profiles, developers can identify source code that may be performance bound. Operators and software testers can also benefit from such performance monitoring process.

Performance profiling presents the time spent by the software application. It breaks the usage into components such as processes, modules and functions. Hotspots are regions of source code that consume significant amount of system resources. Thus, identifying hotspots can help developers locate the regions of code (bottlenecks) that they can work on to optimize the software application.

Linux executables following the Executable and Linkable Format (ELF) contains a symbol table known as "symtab". This section contains the description of each method, including its length and starting instruction address. Thus, it provides sufficient information to translate an instruction pointer to the function. If we also have "Debug info" in ELF, we can translate it to the source code too.

Using statistical sampling, we can estimate the utilizations of each function in the software application. Statistical sampling is also used by other tools such as VTune. To efficiently deploy Bianque into thousands of servers in a data center, we separate it into three roles.

1. Daemon Server – to collect samples and provide storage for the data. It is important this task has low overhead as it is running all the time.

2. Binary Server – to resolve symbols and analyze data. This task is triggered only when a user queries the system. Also this task does not run on the customer facing server. So the overhead can be a higher than that of the daemon server.

3. Builder Server – to store symbol tables in ELF files. This is essentially a storage server keeping tracks of different OS versions and symbol tables running in the servers. It provides storage for both the deamon and the binary servers described above.

# 5   Summary and Discussion

In this paper, we identified challenges to understanding cloud computing performance from the aspects of computing infrastructure, data collection, data analytics and production environment. While solving all these challenges is not currently feasible, understanding these challenges provides us directions in terms of proposing frameworks for cloud computing characterization. An analytics framework is in need to discover the hidden patterns within cloud performance data and thus extract insights for improving software performance in cloud computing. This framework shall utilize time series performance data including user experience data, software performance metrics and system performance data. Needless to say, due to high complexity in cloud computing environment, characterizing the cloud data center becomes fundamentally different from characterizing stand-alone enterprise workloads. Moreover, there are many aspects of cloud computing open to further optimization, and optimizing with respect to one single objective is not realistic. A joint optimization solution for software performance in cloud computing must be sought. When looking into the intersection of cloud computing, performance analysis, mathematical programming and machine learning, we notice a systematic solution for cloud performance enhancement could be born from there. A cloud computing characterization approach that combines performance engineering domain expertise and the rich data attainable from the cloud with advanced data science techniques is essential if we hope to generate robust conclusions.

# Acknowledgments

The authors wish to express their gratitude to their reviewers, Rick Anderson and Sue Bartlett, whose assistance and feedback have been invaluable.

*The 2nd ACM SIGOPS Asia-Pacific Workshop on Systems (APSys 2011)*, Shanghai, China (July 11-12, 2011)